\def\rfr#1{eq. (\ref{#1})}

\def\Rfr#1{Eq. (\ref{#1})}

\def\virg#1{``#1''}

\def\eqi{\begin{equation}}
\def\eqf{\end{equation}}
\def\eqia{\begin{eqnarray}}
\def\eqfa{\end{eqnarray}}
\def\rp#1#2{{#1\over#2}} \def\lb#1{\label{#1}}

\documentclass[11pt]{article}
\usepackage{amsmath,amsthm,amscd,amssymb}
\usepackage{latexsym}
\usepackage{graphicx,epsfig}
\usepackage{lscape,rotating}
\begin{document}

\noindent{\bf \LARGE{Gravitomagnetism and gravitational waves}}
\\
\\
\\
{L. Iorio$^{\ast}$, C. Corda$^{\S}$}\\
{\it $^{\ast}$Fellow of the Royal Astronomical Society. Address for correspondence: Viale Unit$\grave{a}$ di Italia 68
70125 Bari (BA), Italy.  \\ e-mail: lorenzo.iorio@libero.it}\\
{\it $^{\S}$Associazione Scientifica Galileo Galilei,
Via Pier Cironi 16 - 59100 PRATO, Italy\\ e-mail: cordac.galilei@gmail.com}

\vspace{4mm}

\begin{abstract}
After extensively reviewing general relativistic \emph{gravitomagnetism}, both historically and phenomenologically, we review in detail the so-called \emph{magnetic} components of gravitational waves (GWs), which have to be taken into account in the context of the total response functions of interferometers for GWs propagating from arbitrary directions.
Following the more recent approaches of this important issue, the analysis of such \emph{magnetic} components will be reviewed in both of standard General Theory of Relativity (GTR) and Scalar Tensor Gravity. Thus, we show in detail that such a \emph{magnetic} component becomes particularly important in the high-frequency portion of the range of ground based interferometers for GWs which arises from the two different theories of gravity.
Our reviewed results show that if one neglects the \emph{magnetic} contribution to the gravitational field of a GW, approximately 15\% of the potential observable signal could, in principle, be lost.

\end{abstract}


\section{General overview of gravitomagnetism}
The term \virg{gravitomagnetism} \cite{Thorne,Rin,Mash} (GM) commonly indicates the collection of those gravitational phenomena regarding orbiting test particles, precessing gyroscopes, moving clocks and
atoms and propagating electromagnetic waves \cite{Tar,Sciaf} which, in the framework of the Einstein's
General Theory of Relativity \cite{ART} (GTR), arise from non-static distributions of matter and energy. In the weak-field and slow motion approximation, the Einstein field equations \cite{FG} of GTR, which is a highly non-linear Lorentz-covariant tensor theory of gravitation, get
linearized \cite{line}, thus looking like  the Maxwellian equations of electromagntism \cite{treatise}. As a consequence, a \virg{gravitomagnetic}
field $\vec{B}_g$, induced by the off-diagonal components $g_{0i},\ i = 1,2,3$ of the space-time metric
tensor related to mass-energy currents, arises. In particular, far from a localized rotating body with angular
momentum $\vec{S}$ the gravitomagnetic field can be written as \cite{PLA}
\eqi \vec{B}_g(\vec{r}) = \rp{G}{cr^3}\left[\vec{S} -3\left(\vec{S}\cdot\hat{r}\right)\hat{r} \right],\lb{gmfield}\eqf where $G$ is the Newtonian gravitational constant and $c$ is the speed of light in vacuum. It affects, e.g., a test particle moving with velocity $\vec{v}$ with a non-central acceleration \cite{PLA}
\eqi \vec{A}_{\rm GM}=\left(\rp{\vec{v}}{c}\right)\times \vec{B}_g,\eqf which is the cause of two of the most famous and empirically investigated GM effects, as we will see in Section 1.1 and  Section 1.2.
\subsection{Historical overview}\lb{due}
The formal analogies between gravitation and electromagnetism date back to the early days of the Coulomb's force law \cite{Cou1,Cou2,Cou3} between two non-moving pointlike electric charges  (1785). Indeed, it follows faithfully the Newtonian inverse-square force law of gravitation \cite{Newton} between two pointlike masses  (1687) whose state of motion is, instead, irrelevant for its validity within the framework of classical mechanics.
 After the electrodynamical forces between current elements were discovered in 1820-1825 by Amp\`{e}re \cite{ampere}, the situation was reversed. Indeed, in 1870 Holzm\"{u}ller \cite{Holz}, in order to study the motion of a test particle attracted by a fixed center, asked whether Newton's law might not be modified in much the same way as that in which Weber \cite{Web1,Web2} in 1846 had modified Coulomb's law for electric charges in an action-at-a-distance fashion by introducing velocity-dependent terms. He found that the trajectory is no longer closed, but can be described by a slowly precessing ellipse. Later, Tisserand \cite{Tiss1,Tiss2} used this method, yielding a further, \virg{magnetic}-like  component of the Sun's gravitational field acting on the solar system's planets, to attempt-unsuccessfully-to explain the anomalous prograde perihelion precession of Mercury\footnote{It found a natural explanation in 1915 by Einstein \cite{Ein} within his GTR. Note that the static, \virg{gravitoelectric} part of the Sun's field is required for the explanation of such a phenomenon: no mass-energy currents are involved.} of 43.98 arcsec cty$^{-1}$, discovered in 1859 by Le Verrier \cite{Ver}. Also L\'{e}vy \cite{Levy} worked in the same conceptual framework to solve that astronomical problem, but without success.
 The advent  of Maxwell's  electromagnetism \cite{Max}, which is a linear field theory replacing the previous action-at-a-distance theories\footnote{A recent action-at-a-distance gravity theory, based on the scalar velocity-dependent Weber-type potential, is due to Assis \cite{Assi}.}, did not discourage
 further attempts to use analogies of electromagnetic equations to solve gravitational problems. Maxwell himself \cite{Max} considered whether Newtonian gravity could be described by a vector field theory, but he did not succeed because of issues encountered with gravitational energy. A later attempt was due to Heaviside \cite{Hea} in 1894 with his Maxwellian vector field theory of gravity. Among other things, he derived a planetary precession induced by the rotating Sun's GM dipole, but it was too small in magnitude and retrograde with respect to the Mercury's precession observed by Le Verrier.

 Another line of reasoning which yielded to consider gravitational forces induced by moving masses was that connected to the need of explaining the origin of inertia. In particular, the idea that rotating bodies may exert not only the static Newtonian gravitational force but also an additional \virg{frame-dragging} on test particles was probably due to Mach \cite{Mach1}. He speculated that even for relative rotations centrifugal forces arise due to some, unspecified gravitational interaction with the masses of the Earth and of the other celestial bodies \cite{Mach2}, certainly quite larger than the mass of the famous Newton's bucket in relative rotation with respect to the water inside. Later, in 1896 the Friedl\"{a}nder brothers \cite{fried} expressed the conviction  that the properties of inertia and gravitation should be finally derived from a unified law.
 A Maxwell-like theory of gravitation, proposed to explain the origin of inertia in the framework of the Mach's principle, is due to Sciama \cite{Scia}.

 After the birth  of the Einstein's Special Theory of Relativity (STR) in 1905 \cite{Ein05}, the problem of a \virg{magnetic}-type component of the gravitational field of non-static mass distributions was tackled in the framework of the search for a consistent relativistic theory of gravitation \cite{ein}. Indeed, bringing together Newtonian gravitation and Lorentz invariance in a consistent field-theoretic framework necessarily requires the introduction of a \virg{magnetic}-type gravitational field of some form \cite{kha,Bed,Kolb}.

 With a preliminary and still incorrect version of GTR, Einstein and Besso in 1913 \cite{colle} calculated the node precession of planets in the field of the rotating Sun; the figures they obtained for Mercury and Venus were incorrect also because they used a wrong value for the solar mass.
 Soon after GTR was put forth by Einstein (1915) \cite{Ein15}, de Sitter \cite{des} in 1916 used it to preliminarily work out the effects of Sun's rotation on planets' perihelia, although he restricted himself to ecliptic orbits only; his result for Mercury ($-0.01$ arcsec cty$^{-1}$) was too large by one order of magnitude because he assumed a
homogenous and uniformly rotating Sun. In 1918 Thirring \cite{Thirr1} analyzed in a short article the formal analogies between the Maxwell equations and the linearized Einstein equations. Later \cite{Thirr1,Thirr2}, Thirring computed the centrifugal and Coriolis-like gravitomagnetic forces occurring inside  a rotating massive shell. Lense and Thirring\footnote{However, in August 1917 Einstein \cite{coll2} wrote to Thirring that he calculated the Coriolis-type field of the rotating Earth and Sun, and its influence on the orbital elements of planets (and moons). A detailed history of the formulation of the so-called Lense-Thirring effect has recently been outlined by Pfister \cite{Pfi}; according to him, it would be more fair to speak about an Einstein-Thirring-Lense effect.} \cite{LT} in 1918 worked out the gravitomagnetic effects on the orbital motions of test particles outside a slowly rotating mass; in particular, they computed the gravitomagnetic rates of both the satellites of Mars (Phobos and Deimos), and of some of the moons of the giant gaseous planets.
They found for the longitude of the ascending node $\Omega$ a prograde precession, while for the argument of pericenter $\omega$ a retrograde precession occurs
\eqi \dot\Omega_{\rm LT}=\rp{2GS}{c^2 a^3 (1-e^2)^{3/2}},\ \dot\omega_{\rm LT}=-\rp{6GS\cos I}{c^2 a^3 (1-e^2)^{3/2}},\eqf where $a,e,I$ are the semimajor axis, the eccentricity and the inclination of the test particle's orbital plane to the central body's equator, respectively.

Another well known general relativistic gravitomagnetic effect consists of the precession of a gyroscope moving in the field of a slowly rotating body. It was derived in 1959 by Pugh \cite{Pu} and in 1960 by Schiff \cite{Sci1,Sci2,Sci3}.
The possible Machian character of the Schiff effect was discussed by Rindler \cite{Rin2}, and Bondi and Samuel \cite{Reply}.

Cosmological GM and Mach's principle have been recently treated by Schmid \cite{Schmid1,Schmid2}.

Certain subtle issues concerning the gravitomagnetic effects inside a rotating massive shell were solved by Pfister and Braun in 1985 \cite{Pfi2}.

For seeming analogies between Maxwellian electromagnetism and the fully non-linear equations of GTR, see
the works by Matte \cite{Matte} and, more recently, by Costa and Herdeiro \cite{potug}. In this framework,
Pascual-S\'{a}nchez \cite{San} discussed the non-existence of a GM dynamo in the linearized, weak-field and slow-motion approximation of GTR. Tartaglia and Ruggiero \cite{tartrug} investigated the possible occurrence of a GM analog of the Meissner effect using the same approximation for GTR. The non-existence of 
such a phenomenon in gravitation and of other putative GM effects has been demonstrated by Pascual-S\'{a}nchez  \cite{San2}. 
\subsection{Experimental/observational overview}\lb{tre}
 About empirical investigations of possible gravitational effects due to moving bodies, in 1896 I. Friedl\"{a}nder \cite{fried} was the first to perform an experiment concerning a putative gravitational influence of moving bodies. He used as a source a rapidly rotating heavy fly wheel and tried-unsuccessfully-to detect its gravitational effects on a torsion balance mounted above the fly wheel, in line with its axis. Later, in 1904 F\"{o}ppl \cite{foppl} looked for possible Coriolis-like gravitational dragging effects induced  on a gyroscope made of two heavy fly wheels by the whole rotating Earth as a source.

 Moving to more recent epochs, soon after the dawn of the space age with the launch of Sputnik in 1957 it was proposed by Soviet scientists to directly test the general relativistic Lense-Thirring effect with artificial satellites orbiting the Earth. In particular, V.L. Ginzburg \cite{Ginz0,Ginz1,Ginz2} proposed to use the perigee of a terrestrial spacecraft in highly elliptic orbit, while  A.F. Bogorodskii \cite{Bogo} considered also the node. In 1959 Yilmaz  \cite{Yil}, aware of the aliasing effect of the much larger classical precessions induced by the non-sphericity of the Earth, proposed to launch a satellite in a polar orbit to cancel them. About twenty years later, in 1976 van Patten and Everitt \cite{Lipa1,Lipa2} suggested to use a pair of drag-free, counter-orbiting terrestrial spacecraft in nearly polar orbits to detect their combined Lense-Thirring node precessions. In 1977-1978 Cugusi and Proverbio \cite{Cugu1,Cugu2} suggested to use the passive geodetic satellite LAGEOS, in orbit around the Earth since 1976 and tracked with the Satellite Laser Ranging (SLR) technique, along with the other existing laser-ranged targets to measure the Lense-Thirring node precession. In 1986 Ciufolini \cite{Ciu86} proposed a somewhat simpler version of the van Patten-Everitt mission consisting of looking at the sum of the nodes of LAGEOS and of another SLR satellite to be launched in the same orbit, apart from the inclination which should be switched by 180 deg in order to minimize the competing classical precessions due to the centrifugal oblateness of the Earth. Iorio \cite{IorioPL} showed that such an orbital configuration would allow, in principle, to use the difference of the perigees as well. Tests have started to be effectively performed later by Ciufolini and coworkers \cite{Ciuetal96} with  the
LAGEOS and LAGEOS II satellites\footnote{LAGEOS II was launched in 1992, but its orbital configuration is different from that proposed in \cite{Ciu86}.}, according to a strategy by
Ciufolini \cite{Ciu96} involving the use of a suitable linear combination of the nodes $\Omega$ of both
satellites and the perigee $\omega$ of LAGEOS II in order to remove the impact of the first two multipoles of the non-spherical gravitational potential of the Earth. Latest tests have been reported by Ciufolini and Pavlis \cite{Ciu04,CiuPeron}, Lucchesi \cite{Luc07} and  Ries and coworkers \cite{Riesetal} with only the nodes of both the satellites according to a combination of them explicitly proposed by Iorio \cite{Iornodi}. The total uncertainty reached is still matter of debate \cite{crit1,crit2,crit3,crit4,crit5,IorioSSRV,IorioCEJP} because of the lingering uncertainties in the Earth's multipoles and in how to evaluate their biasing impact; it may be as large as $\approx 20-30\%$ according to conservative evaluations \cite{crit1,crit4,crit5,IorioSSRV,IorioCEJP}, while more optimistic views \cite{Ciu04,CiuPeron,Riesetal} point towards $10-15\%$.
Several authors \cite{multi1,multi2,multi3,multi4,multi5} explored the possibility of using other currently orbiting SLR geodetic satellites in addition to LAGEOS and LAGEOS II.
A new SLR geodetic satellite, named LARES, should be launched by the Italian Space Agency (ASI) in 2010 to improve the present-day accuracy of the Lense-Thirring tests by combining its node with those of the existing LAGEOS and LAGEOS II \cite{IorioNA}. The claimed accuracy is  $1\%$ \cite{Ciufo}, but also in this case the realistic level of uncertainty may be quite larger \cite{crit6,crit7,IorioASR,IorioGRG} because of the relatively low orbit of LARES with respect to LAGEOS and LAGEOS II which should bring into play the systematic alias by several non-perfectly known Earth's multipoles.

 In 2006 a preliminary test in the gravitational field of Mars with the Mars Global Surveyor (MGS) has been performed by Iorio \cite{IorioMGS1,IorioMGS2}. He interpreted certain features of the time series of the out-of-plane portion $N$ of the MGS orbit involving its node in terms of the Lense-Thirring effect. In particular, the average of the Root-Mean-Square
(RMS) orbit-overlap differences of the out-of-plane component of the MGS path over 5 years agrees with the predicted average Lense-Thirring out-of-plane shift over the same time span within a few percent, while a linear fit to the complete $N$ time series for the entire MGS
data set shows an agreement with the corresponding predicted Lense-Thirring signal at a $\thicksim 40\%$. A debate about the validity of such an interpretation arose \cite{Krogh}, and it is still ongoing \cite{IorioMGS3}.
The possibility of designing a dedicated mission to Mars has been recently considered by Iorio \cite{IorioMARS}.

Iorio and Lainey \cite{IorJUP} revisited the original proposal by Lense and Thirring \cite{LT} concerning the system of Jovian moons in view of recent advances in orbit determination of the four large Galilean satellites. The possibility of using fruitfully them seem still to be premature.

Concerning the Sun's GM field and the inner planets, the situation is nowadays more favorable than in the past \cite{des,Cugu2}. Indeed,
the astronomer Pitjeva \cite{Pit1,Pit2} has recently fitted the full set of dynamical force models of the planetary motions of the EPM ephemerides to about one century of data of several types by estimating, among other things, corrections $\Delta\dot\varpi$ to the standard Newtonian/Einsteinian secular precessions of the longitude of the perihelia of all the rocky planets.
In doing so she did not model the solar GM field, so that such corrections, by construction, are well suited to test the Lense-Thirring effect \cite{IorAA,PSS,IorSRX}.
The magnitude of the predicted Lense-Thirring perihelion precessions, although one order of magnitude smaller than what argued in  earlier studies \cite{des,Cugu2}, is about of the same order of magnitude of, or even larger than,  the present-day uncertainty in the estimated $\Delta\dot\varpi$. In particular, it has been noted \cite{IorSRX} that the Lense-Thirring perihelion precession of Venus amounts to $\dot\varpi_{\rm LT} = -0.0003$ arcsec cty$^{-1}$, while the estimated correction for Venus is
$\Delta\dot\varpi=-0.0004\pm 0.0001$, where the quoted uncertainty is the $1-\sigma$ statistical error. Thus, the existence of the Lense-Thirring effect would be confirmed at a $25\%$ level, although caution is in order because the realistic uncertainty in $\Delta\dot\varpi$ might be up to 5 times larger. The systematic bias due to the mismodelling of other competing classical effects would be less relevant than in the LAGEOS-LAGEOS II case.
The proposed space-based Astrodynamical Space Test of Relativity using Optical Devices (ASTROD) mission \cite{astrod} has, among its scientific goals, also the accurate determination of the Sun's angular momentum through a $\simeq 1\%$ measurement of the gravitomagnetic time delay on electromagnetic waves.

Soon after the formulation of the Schiff effect,  in 1961 Fairbank and Schiff \cite{Fair} submitted to NASA a proposal for a dedicated space-based experiment aimed to directly measure it. Such an extremely complicated  mission, later named Gravity Probe B (GP-B) \cite{gpb1,gpb2}, consisted of a drag-free, liquid helium-cooled spacecraft moving in a polar, low orbit around the Earth and carrying onboard four superconducting gyroscopes whose GM precessions should have been detected by Superconducting Quantum Interference Devices (SQUID) with an expected accuracy of $1\%$ or better. It took 43 years to be implemented since GP-B was finally launched on 20 April 2004; the science data collection  lasted from 27 August 2004 to 29 September 2005, while the data analysis is still ongoing \cite{ANA1,ANA2}. It seems that the final accuracy obtainable will be not so good as initially hoped because of the occurrence of unexpected systematic errors \cite{merdacce1,merdacce2,merdacce3}.
At present, GP-B team reports\footnote{See on the WEB: http://einstein.stanford.edu/} a statistical error of $\thicksim 14\%$ and systematic uncertainty of $\thicksim
10\%$.
In 1975 Haas and Ross \cite{haa} proposed to measure the angular momenta of the Sun and Jupiter by exploiting the Schiff effect with dedicated spacecraft-based missions, but such a proposal was not carried out so far.

All the previously reviewed attempts aim to obtain direct tests of some GM effects. However, according to Nordtvedt \cite{Nord1,Nord2}, the GM interaction would have already been observed, with a relative accuracy of 1 part to 1000, in comprehensive fits of the motions of  several astronomical and astrophysical bodies like satellites, binary pulsars and the Moon. In fact, Nordtvedt does not refer to the effects considered so far, caused by the rotation of the body which acts as source of the gravitational field (\virg{intrinsic} GM). Instead, he primarily deals  with some GM long-periodic harmonic perturbations affecting the Earth-Moon range induced by the translational GM mass currents due to the orbital motion of the Earth-Moon system around the Sun (\virg{extrinsic} GM); Lunar Laser Ranging (LLR) would have measured them with a $0.1\%$ accuracy \cite{Nord3}. However, Kopeikin argues that LLR would not be able to detect genuine GM signatures which are not spurious, gauge-dependent effects \cite{Kop1}. For other works about such an issue, see \cite{llr1,llr2,llr3}.  Concerning the possibility of directly measuring the Lense-Thirring precessions of the Moon's motion due to the Earth's angular momentum, it has been recently proven to be still unfeasible by Iorio \cite{IorioLLR} because of the  too large level of uncertainty in several competing classical effects. A test of
extrinsic gravitomagnetism concerning the deflection of electromagnetic waves by
Jupiter in its orbital motion has been reported in a dedicated analysis of radiointerferometric
data by Fomalont and Kopeikin \cite{Kop3}, but also such a test is controversial: see the WEB page by Will at http://physics.wustl.edu/cmw/SpeedofGravity.html.

For other proposals to directly detect various aspects of the (intrinsic) GM field in Earth-based laboratory and space-based experiments, see, e.g., the book by Iorio \cite{Iorbook}.  Extensive overviews of the importance of GM in astrophysical contexts like accretion disks around compact objects and relativistic jets in quasars and galactic nuclei can be found, e.g., in the book by Thorne, Price and MacDonald \cite{membra}, and in Section E of the book by Ruffini and Sigismondi \cite{Ruff}.

\section{\emph{Magnetic} component in the gravitational field of a gravitational wave}

\subsection{The importance of gravitational waves: a new window into the Universe}\lb{tre}
The data analysis of interferometric Gravitational Waves (GWs) detectors has nowadays been started,
 and the scientific community hopes in a first direct detection of GWs in next years; for the current status of GWs interferometers see Ref.~\cite{Giazotto}. In such a way, the indirect evidence of the existence of GWs by Hulse and Taylor \cite{Pulsar}, Nobel Prize winners, will be confirmed.
Detectors for GWs will be important for a better knowledge of the Universe \cite{Giazotto} and also because the interferometric GWs detection will be the definitive test for GTR or, alternatively, a strong endorsement for Extended Theories of Gravity \cite{Essay}. In fact, if advanced projects on the detection of GWs improve their sensitivity, allowing the Scientific Community to perform a GW astronomy, accurate angle- and frequency-dependent response functions of interferometers for GWs arising from various theories of gravity will permit to discriminate among GTR and extended theories of gravity. This  ultimate test will work because standard GTR admits only two polarizations for GWs, while in all extended theories the polarizations are, at least, three, see \cite{Essay} for details.

On the other hand,  the discovery of GW emission by the compact binary system composed by two Neutron Stars PSR1913+16 \cite{Pulsar}  has been, for  physicists working in this field, the ultimate thrust allowing to reach the extremely sophisticated technology needed for investigating in this field of research \cite{Giazotto}.

Gravitational Waves are a consequence of Einstein's GTR \cite{Albertino}, which presuppose GWs to be ripples in the space-time curvature travelling at light speed \cite{AlbertinoII,AlbertinoIII}. Only asymmetric astrophysics sources can emit GWs. The most efficient are coalescing binaries systems, while a single rotating pulsar can rely only on spherical asymmetries, usually very small. Supernovae could have relevant asymmetries, being potential sources \cite{Giazotto}.

The most important cosmological source of GWs is, in principle, the so called  stochastic background of GWs which, together with the Cosmic Background Radiation (CBR), would carry, if detected, a huge amount of information on the early stages of the Universe evolution \cite{rainbow, rainbowII, rainbowIII}.
The existence of a relic stochastic background of GWs is a consequence of generals assumptions. Essentially it derives from a mixing between
basic principles of classical theories of gravity and of quantum field theory. The strong variations of the gravitational field in the early
universe amplify the zero-point quantum oscillations and produce relic GWs. It is well known that the detection of relic GWs is the only way to learn about the evolution of the very early universe, up to the bounds of the Planck epoch and the initial singularity \cite{Allen, rainbowIV}.
It is very important to stress the unavoidable and fundamental character of this mechanism. The model derives from the inflationary scenario for the early universe \cite{Inflation}, which is tuned in a good way with the WMAP data on the CBR (in particular exponential inflation and spectral index $\approx1$ \cite{Cosmology}). Inflationary models  are cosmological models in which the Universe undergoes a brief phase of a very rapid expansion in early times \cite{Inflation}. In this context the expansion could be power-law or exponential in time. Such models provide solutions to the horizon and flatness problems and contain a mechanism which creates perturbations in all fields \cite{Allen, rainbowIV, Inflation}. Important for our case
is that this mechanism also provides a distinctive spectrum of relic GWs \cite{Allen, rainbowIV}. The GWs perturbations arise from the uncertainty principle and the spectrum of relic GWs is generated from the adiabatically-amplified zero-point fluctuations \cite{Allen, rainbowIV}.

Regarding the potential GW detection, let us recall some historical notes.

In 1957,  F.A.E. Pirani, who was a member of the Bondi's research group, proposed the geodesic deviation equation as a tool for designing a practical GW detector \cite{Pirani}.

In 1959, Joseph Weber studied a detector that, in principle, might be able to measure displacements smaller than the size of the nucleus \cite{Weber}. He developed an experiment using a large suspended bar of aluminum, with a high resonant Q at a frequency of about 1 kHz. Then, in 1960, he tried to test the general relativistic prediction of gravitational waves from strong gravity collisions \cite{Weber2} and, in 1969, he claimed evidence for observation of gravitational waves (based on coincident signals)  from two bars separated by 1000 km \cite{web}. He also proposed the idea of doing an experiment to detect gravitational waves using laser interferometers \cite{web}. In fact, all the modern detectors can be considered like being originated from early Weber's ideas \cite{Giazotto}.

At the present time, in the world there are  five cryogenic bar detectors have been built to work at very low temperatures ($<4K$): Explorer at CERN, Nautilus at Frascati INFN National Laboratory, Auriga at Legnaro National Laboratory, Allegro at Luisiana State University  and Niobe in Perth \cite{Giazotto}. Instrumental details can be found in \cite{Giazotto} and references within.
Spherical detectors are the Mario Schenberg, which has been built in San Paolo (Brazil) and the MiniGRAIL, which has been built at the Kamerlingh Onnes Laboratory of Leiden University, see \cite{Giazotto} and references within. Spherical detectors are important for the potential detection of the \emph{scalar} component of GWs that is admitted by Extended Theories of Gravity \cite{correlation}.
In the case of interferometric detectors, free falling masses are interferometer mirrors which can be separated by kilometres (3km for Virgo, 4km for LIGO). In this way,  GW tidal force is, in principle, several order of magnitude larger than in bar detectors. Interferometers have very large bandwidth (10-10000 Hz) because mirrors are suspended to pendulums having resonance in the Hz region. Thus, above such a resonance frequency, mirrors works, in a good approximation,  like freely falling masses in the horizontal plane \cite{Giazotto}.

Recently, starting from the analysis in Ref.~\cite{Grishchuk}, some papers in  literature have shown the importance of the gravitomagnetic effects in the framework of the GWs detection too \cite{Corda,Corda2,Corda3}.
In fact, the so-called \emph{magnetic} components of GWs have to be taken into account in the context of the total response functions of interferometers for GWs propagating from arbitrary directions, \cite{Grishchuk, Corda, Corda2, Corda3}. In next analysis we will show that such a \emph{magnetic} component becomes particularly important in the high-frequency portion of the range of ground based interferometers for GWs which arises from standard GTR.

In a recent paper, the \emph{magnetic}  component has been extended to GWs arising from scalar-tensor gravity too \cite{Cafaro}. In particular, in Ref.~\cite{Cafaro} it has been shown that if one neglects the \emph{magnetic} contribution considering only the low-frequency approximation of the \emph{electric} contribution, a portion of about the $15\%$ of the signal could be, in principle, lost in the case of Scalar Tensor Gravity too, in total analogy with the standard case of GTR \cite{Grishchuk, Corda, Corda2, Corda3}.

For the sake of completeness, such a case will be included in the following discussion.

\subsection{The \emph{magnetic} component of GWs in standard GTR}\lb{tre}
In a laboratory environment on Earth  coordinate systems in which the space-time is locally flat are typically used, and the distance between any two points is given simply by the difference in their coordinates in the sense of Newtonian physics \cite{Cafaro, Gravitation, Landau}.
In this frame, called the frame of the local observer, GWs manifest them-self by exerting tidal forces on the masses (the mirror and the
beam-splitter in the case of an interferometer \cite{Corda, Corda2, Corda3, Cafaro}, see Figure 1.
\begin{figure}
\includegraphics{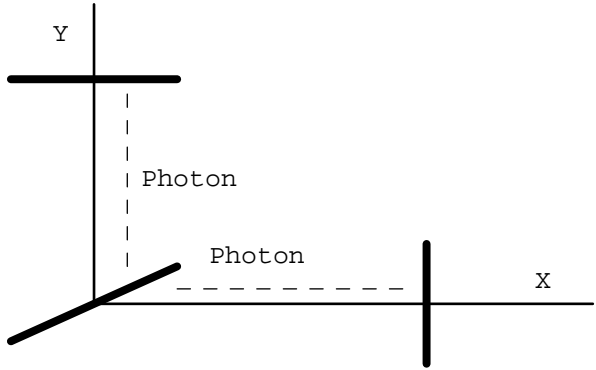}\lb{figurauno}

\caption{photons can be launched from the beam-splitter to be bounced back
by the mirror}

\end{figure}

The presence and importance of the so-called \emph{magnetic} components of GWs in the framework of GTR was emphasized by Baskaran and Grishchuk that computed the correspondent detector patterns \cite{Grishchuk},   while  more detailed angular and frequency dependences of the response functions for the \emph{magnetic} components have been given, with a specific application to the parameters of the LIGO and Virgo interferometers, in Refs.~\cite{Corda, Corda2, Corda3}.
Thus, following Refs.~\cite{Corda, Corda2, Corda3}, we will, now, work with $G=1$, $c=1$ and $\hbar=1$ and can call $h_{+}(t_{tt}+z_{tt})$ and $h_{\times}(t_{tt}+z_{tt})$ the weak perturbations due to the $+$ and the $\times$ polarizations which are expressed in terms of synchronous coordinates $t_{tt},x_{tt},y_{tt},z_{tt}$
in the transverse-traceless (TT) gauge. In this way, the most general GW propagating in the $z_{tt}$ direction can be written in terms of a plane monochromatic wave \cite{Corda, Corda2, Corda3}
\begin{equation}
\begin{array}{c}
h_{\mu\nu}(t_{tt}+z_{tt})=h_{+}(t_{tt}+z_{tt})e_{\mu\nu}^{(+)}+h_{\times}(t_{tt}+z_{tt})e_{\mu\nu}^{(\times)}=\\
\\=h_{+0}\exp i\omega(t_{tt}+z_{tt})e_{\mu\nu}^{(+)}+h_{\times0}\exp i\omega(t_{tt}+z_{tt})e_{\mu\nu}^{(\times)},\end{array}\label{eq: onda generale}\end{equation}
and the corresponding line element will be
\begin{equation}
ds^{2}=dt_{tt}^{2}-dz_{tt}^{2}-(1+h_{+})dx_{tt}^{2}-(1-h_{+})dy_{tt}^{2}-2h_{\times}dx_{tt}dx_{tt}.\label{eq: metrica TT totale}\end{equation}
The wordlines $x_{tt},y_{tt},z_{tt}=const.$ are timelike geodesics representing the histories of free test masses \cite{Grishchuk,Corda, Corda2, Corda3}.
The coordinate transformation $x^{\alpha}=x^{\alpha}(x_{tt}^{\beta})$ from the TT coordinates to the frame of the local observer is \cite{Grishchuk,Corda, Corda2, Corda3}
\begin{equation}
\begin{array}{c}
t=t_{tt}+\frac{1}{4}(x_{tt}^{2}-y_{tt}^{2})\dot{h}_{+}-\frac{1}{2}x_{tt}y_{tt}\dot{h}_{\times}\\
\\x=x_{tt}+\frac{1}{2}x_{tt}h_{+}-\frac{1}{2}y_{tt}h_{\times}+\frac{1}{2}x_{tt}z_{tt}\dot{h}_{+}-\frac{1}{2}y_{tt}z_{tt}\dot{h}_{\times}\\
\\y=y_{tt}+\frac{1}{2}y_{tt}h_{+}-\frac{1}{2}x_{tt}h_{\times}+\frac{1}{2}y_{tt}z_{tt}\dot{h}_{+}-\frac{1}{2}x_{tt}z_{tt}\dot{h}_{\times}\\
\\z=z_{tt}-\frac{1}{4}(x_{tt}^{2}-y_{tt}^{2})\dot{h}_{+}+\frac{1}{2}x_{tt}y_{tt}\dot{h}_{\times},\end{array}\label{eq: trasf. coord.}\end{equation}
where it is $\dot{h}_{+}\doteq\frac{\partial h_{+}}{\partial t}$ and $\dot{h}_{\times}\doteq\frac{\partial h_{\times}}{\partial t}$.
The coefficients of this transformation (components of the metric and its first time derivative) are taken along the central wordline of the local observer \cite{Grishchuk,Corda, Corda2, Corda3}.
It is well known from Refs.~\cite{Grishchuk,Corda, Corda2, Corda3} that the linear and quadratic terms, as powers of $x_{tt}^{\alpha}$, are unambiguously
determined by the conditions of the frame of the local observer, while the cubic and higher-order corrections are not determined by these
conditions. Thus, at high-frequencies, the expansion in terms of higher-order corrections breaks down \cite{Grishchuk,Corda, Corda2, Corda3}.

Considering a free mass riding on a timelike geodesic ($x=l_{1}$, $y=l_{2},$ $z=l_{3}$) \cite{Grishchuk,Corda, Corda2, Corda3},
\rfr{eq: trasf. coord.} defines the motion of this mass with respect to the introduced frame of the local observer. In concrete terms one gets
\begin{equation}
\begin{array}{c}
x(t)=l_{1}+\frac{1}{2}[l_{1}h_{+}(t)-l_{2}h_{\times}(t)]+\frac{1}{2}l_{1}l_{3}\dot{h}_{+}(t)+\frac{1}{2}l_{2}l_{3}\dot{h}_{\times}(t)\\
\\y(t)=l_{2}-\frac{1}{2}[l_{2}h_{+}(t)+l_{1}h_{\times}(t)]-\frac{1}{2}l_{2}l_{3}\dot{h}_{+}(t)+\frac{1}{2}l_{1}l_{3}\dot{h}_{\times}(t)\\
\\z(t)=l_{3}-\frac{1}{4[}(l_{1}^{2}-l_{2}^{2})\dot{h}_{+}(t)+2l_{1}l_{2}\dot{h}_{\times}(t),\end{array}\label{eq: Grishuk 0}\end{equation}
which are exactly eqs. (13) of Ref.~\cite{Grishchuk} rewritten using the notation of Refs.~\cite{Corda, Corda2, Corda3}.
In absence of GWs, the position of the mass is $(l_{1},l_{2},l_{3}).$
The effect of the GW is to drive the mass to have oscillations. Thus, in general, from \rfr{eq: Grishuk 0} all three components of motion are present \cite{Grishchuk,Corda, Corda2, Corda3}.
Neglecting the terms with $\dot{h}_{+}$ and $\dot{h}_{\times}$ in \rfr{eq: Grishuk 0}, the \emph{traditional} equations for the mass motion are obtained \cite{Grishchuk,Corda, Corda2, Corda3}
\begin{equation} \begin{array}{c}
x(t)=l_{1}+\frac{1}{2}[l_{1}h_{+}(t)-l_{2}h_{\times}(t)]\\
\\y(t)=l_{2}-\frac{1}{2}[l_{2}h_{+}(t)+l_{1}h_{\times}(t)]\\
\\z(t)=l_{3}.\end{array}\label{eq: traditional}\end{equation}
Clearly, this is the analogous of the \emph{electric} component of motion in electrodynamics \cite{Grishchuk,Corda, Corda2, Corda3}, while equations
\begin{equation}
\begin{array}{c}
x(t)=l_{1}+\frac{1}{2}l_{1}l_{3}\dot{h}_{+}(t)+\frac{1}{2}l_{2}l_{3}\dot{h}_{\times}(t)\\
\\y(t)=l_{2}-\frac{1}{2}l_{2}l_{3}\dot{h}_{+}(t)+\frac{1}{2}l_{1}l_{3}\dot{h}_{\times}(t)\\
\\z(t)=l_{3}-\frac{1}{4[}(l_{1}^{2}-l_{2}^{2})\dot{h}_{+}(t)+2l_{1}l_{2}\dot{h}_{\times}(t),\end{array}\label{eq: news}\end{equation}
are the analogous of the \emph{magnetic} component of motion.
One could think that the presence of these \emph{magnetic} components is a \emph{frame artefact} due to the transformation \rfr{eq: trasf. coord.}, but in Section 4 of Ref.~\cite{Grishchuk} \rfr{eq: Grishuk 0} have been directly obtained from the geodesic deviation equation too, thus the magnetic components have a real physical significance.
The fundamental point of Ref.~\cite{Grishchuk,Corda, Corda2, Corda3} is that the \emph{magnetic} components become important when the frequency of the wave increases  but only in the low-frequency regime. This can be understood directly from \rfr{eq: Grishuk 0}. In fact, using \rfr{eq: onda generale} and \rfr{eq: trasf. coord.}, \rfr{eq: Grishuk 0} become
\begin{equation}
\begin{array}{c}
x(t)=l_{1}+\frac{1}{2}[l_{1}h_{+}(t)-l_{2}h_{\times}(t)]+\frac{1}{2}l_{1}l_{3}\omega h_{+}(t-\frac{\pi}{2})+\frac{1}{2}l_{2}l_{3}\omega h_{\times}(t-\frac{\pi}{2})\\
\\y(t)=l_{2}-\frac{1}{2}[l_{2}h_{+}(t)+l_{1}h_{\times}(t)]-\frac{1}{2}l_{2}l_{3}\omega h_{+}(t-\frac{\pi}{2})+\frac{1}{2}l_{1}l_{3}\omega h_{\times}(t-\frac{\pi}{2})\\
\\z(t)=l_{3}-\frac{1}{4[}(l_{1}^{2}-l_{2}^{2})\omega h_{+}(t-\frac{\pi}{2})+2l_{1}l_{2}\omega h_{\times}(t-\frac{\pi}{2}).\end{array}\label{eq: Grishuk 01}\end{equation}
Thus, the terms with $\dot{h}_{+}$ and $\dot{h}_{\times}$ in  \rfr{eq: Grishuk 0} can be neglected only when the wavelength goes to infinity \cite{Grishchuk,Corda, Corda2, Corda3}, while, at high-frequencies, the expansion in terms of $\omega l_{i}l_{j}$ corrections, with $i,j=1,2,3,$ breaks down \cite{Grishchuk,Corda, Corda2, Corda3}.

Now, let us compute the total response functions of interferometers for the \emph{magnetic} components in standard GTR.

Equations \rfr{eq: Grishuk 0}, that represent the coordinates of the mirror of the interferometer in presence of a GW in the frame of the local observer, can be rewritten for the pure magnetic component of the $+$ polarization as

\begin{equation}
\begin{array}{c}
x(t)=l_{1}+\frac{1}{2}l_{1}l_{3}\dot{h}_{+}(t)\\
\\y(t)=l_{2}-\frac{1}{2}l_{2}l_{3}\dot{h}_{+}(t)\\
\\z(t)=l_{3}-\frac{1}{4}(l_{1}^{2}-l_{2}^{2})\dot{h}_{+}(t),\end{array}\label{eq: Grishuk 1}\end{equation}

where $l_{1},l_{2}\textrm{ }and\textrm{ }\textrm{ }l_{3}$ are the unperturbed coordinates of the mirror.

To compute the response functions for an arbitrary propagating direction of the GW, one recalls that the arms of the interferometer are in general
in the $\overrightarrow{u}$ and $\overrightarrow{v}$ directions, while the $x,y,z$ frame is adapted to the propagating GW (i.e. the observer is assumed located in the position of the beam splitter). Then, a spatial rotation of the coordinate system has to be performed:

\begin{equation}
\begin{array}{ccc}
u & = & -x\cos\theta\cos\phi+y\sin\phi+z\sin\theta\cos\phi\\
\\v & = & -x\cos\theta\sin\phi-y\cos\phi+z\sin\theta\sin\phi\\
\\w & = & x\sin\theta+z\cos\theta,\end{array}\label{eq: rotazione magn}\end{equation}

or, in terms of the $x,y,z$ frame:

\begin{equation}
\begin{array}{ccc}
x & = & -u\cos\theta\cos\phi-v\cos\theta\sin\phi+w\sin\theta\\
\\y & = & u\sin\phi-v\cos\phi\\
\\z & = & u\sin\theta\cos\phi+v\sin\theta\sin\phi+w\cos\theta.\end{array}\label{eq: rotazione 2 magn}\end{equation}

In this way, the GW is propagating from an arbitrary direction $\overrightarrow{r}$ to the interferometer (see Figure 2).
\begin{figure}
\includegraphics{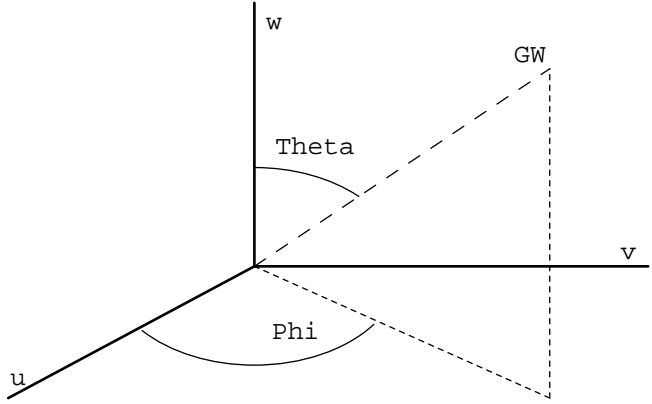}\lb{figuradue}

\caption{a GW propagating from an arbitrary direction}

\end{figure}
As the mirror of   \rfr{eq: Grishuk 1} is situated in the $u$ direction, using   \rfr{eq: Grishuk 1}, \rfr{eq: rotazione magn} and \rfr{eq: rotazione 2 magn} the $u$ coordinate of the mirror is given by

\begin{equation}
u=L+\frac{1}{4}L^{2}A\dot{h}_{+}(t),\label{eq: du magn}\end{equation}

where

\begin{equation}
A\doteq\sin\theta\cos\phi(\cos^{2}\theta\cos^{2}\phi-\sin^{2}\phi)\label{eq: A}\end{equation}

and $L=\sqrt{l_{1}^{2}+l_{2}^{2}+l_{3}^{2}}$ is the length of the interferometer arms.

The computation for the $v$ arm is similar to the one above. Using   \rfr{eq: Grishuk 1}, \rfr{eq: rotazione magn} and \rfr{eq: rotazione 2 magn},
the coordinate of the mirror in the $v$ arm is

\begin{equation}
v=L+\frac{1}{4}L^{2}B\dot{h}_{+}(t),\label{eq: dv magn}\end{equation}

where
\begin{equation}
B\doteq\sin\theta\sin\phi(\cos^{2}\theta\cos^{2}\phi-\sin^{2}\phi).\label{eq: B}\end{equation}

\Rfr{eq: du magn} and \rfr{eq: dv magn} represent the distance of the two mirrors of the interferometer from the beam-splitter in presence of the GW (note that only the contribution of the \emph{magnetic} component of the $+$ polarization of the GW is taken into account).
They represent particular cases of the more general form given in eq. (33) of \cite{Grishchuk}.

A \emph{signal} can also be defined in the time domain ($T=L$ in our notation)
\begin{equation}
\frac{\delta T(t)}{T}\doteq\frac{u-v}{L}=\frac{1}{4}L(A-B)\dot{h}_{+}(t).\label{eq: signal piu}\end{equation}

The quantity \rfr{eq: signal piu} can be computed in the frequency domain by using the Fourier transform of $h_{+}$, defined by

\begin{equation}
\tilde{h}_{+}(\omega)=\int_{-\infty}^{\infty}dth_{+}(t)\exp(i\omega t),\label{eq: trasformata di fourier magn}\end{equation}
obtaining

\[\frac{\tilde{\delta}T(\omega)}{T}=H_{\rm magn}^{+}(\omega)\tilde{h}_{+}(\omega),\]

where the function

\begin{equation}
\begin{array}{c}
H_{\rm magn}^{+}(\omega)=-\frac{1}{8}i\omega L(A-B)=\\
\\=-\frac{1}{4}i\omega L\sin\theta[(\cos^{2}\theta+\sin2\phi\frac{1+\cos^{2}\theta}{2})](\cos\phi-\sin\phi)\end{array}\label{eq: risposta totale magn}\end{equation}

is the total response function of the interferometer for the \emph{magnetic} component of the $+$ polarization \cite{Grishchuk,Corda, Corda2, Corda3}.

In the above computation the theorem on the derivative of the Fourier transform has been used.

The angular dependence of the response function \rfr{eq: risposta totale magn} of the  LIGO interferometer to the magnetic component of the $+$ polarization for $f=8000$ Hz is shown in Figure 3.

\begin{figure}
\includegraphics{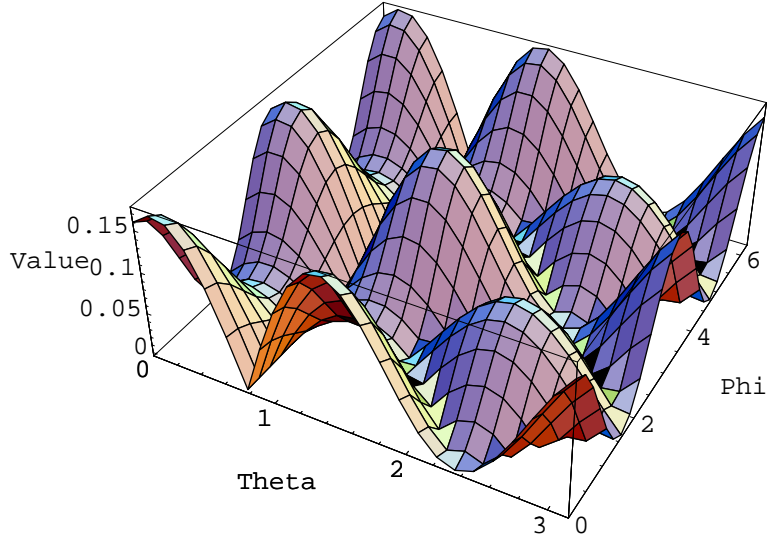}\lb{figuratre}

\caption{the angular dependence of the response function of the LIGO interferometer
to the magnetic component of the $+$ polarization for $f=8000$ Hz}

\end{figure}

The analysis can be generalized for the \emph{magnetic} component of the $\times$ polarization too. In this case,  \rfr{eq: Grishuk 0} can be rewritten for the pure magnetic component of the $\times$ polarization as \cite{Grishchuk,Corda, Corda2, Corda3}
\begin{equation}
\begin{array}{c}
x(t)=l_{1}+\frac{1}{2}l_{2}l_{3}\dot{h}_{\times}(t)\\
\\y(t)=l_{2}+\frac{1}{2}l_{1}l_{3}\dot{h}_{\times}(t)\\
\\z(t)=l_{3}-\frac{1}{2}l_{1}l_{2}\dot{h}_{\times}(t).\end{array}\label{eq: Grishuk 2}\end{equation}
Using   \rfr{eq: Grishuk 2}, \rfr{eq: rotazione magn} and \rfr{eq: rotazione 2 magn}, the $u$ coordinate of the mirror in the $u$ arm of the interferometer is given by
\begin{equation} u=L+\frac{1}{4}L^{2}C\dot{h}_{\times}(t),\label{eq: du C}\end{equation}
where
\begin{equation}
C\doteq-2\cos\theta\cos^{2}\phi\sin\theta\sin\phi,\label{eq: C}\end{equation}
while the $v$ coordinate of the mirror in the $v$ arm of the interferometer
is given by \begin{equation}
v=L+\frac{1}{4}L^{2}D\dot{h}_{\times}(t),\label{eq: dv  D}\end{equation}
with
\begin{equation}
D\doteq2\cos\theta\cos\phi\sin\theta\sin^{2}\phi.\label{eq: D}\end{equation}
Thus, with an analysis similar to the one of previous Sections, it is possible to show that the response function of the interferometer for the magnetic component of the $\times$ polarization is \cite{Grishchuk,Corda, Corda2, Corda3}
\begin{equation}
\begin{array}{c}
H_{\rm magn}^{\times}(\omega)=-i\omega T(C-D)=\\
\\=-i\omega L\sin2\phi(\cos\phi+\sin\phi)\cos\theta,\end{array}\label{eq: risposta totale 2 per magn}\end{equation}

The angular dependence of the response function \rfr{eq: risposta totale 2 per magn} of the  LIGO interferometer to the magnetic component of the $\times$ polarization for $f=8000$ Hz is shown in Figure 4.

\begin{figure}
\includegraphics{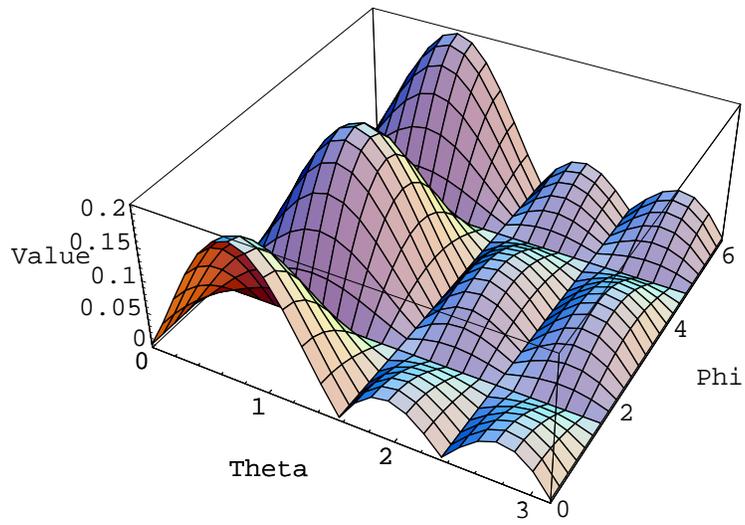}\lb{figuraquattro}

\caption{the angular dependence of the total response function of the LIGO
interferometer to the magnetic component of the $\times$ polarization for $f=8000$ Hz}

\end{figure}

From Figure 3 and Figure 4, it looks clear that if one neglects the \emph{magnetic} contribution, approximately 15\% of currently observable signal could,
in principle, be lost.

\subsection{The \emph{magnetic} component of GWs in Scalar Tensor Gravity}\lb{tre}

In the framework of Scalar Tensor Gravity, the TT gauge can be extended to a third polarization \cite{Essay, Cafaro, Lee, Damour, Tobar, Nakao, Capozziello}.
In this way,  the total perturbation of a gravitational wave propagating in the $z-$ direction in this gauge is \cite{Essay, Cafaro, Capozziello}

\begin{equation}
h_{\mu\nu}(t+z)=A^{+}(t+z)e_{\mu\nu}^{(+)}+A^{\times}(t+z)e_{\mu\nu}^{(\times)}+\Phi(t+z)e_{\mu\nu}^{(s)}.\label{eq: perturbazione totale}\end{equation}

The term $A^{+}(t+z)e_{\mu\nu}^{(+)}+A^{\times}(t+z)e_{\mu\nu}^{(\times)}$ describes the two standard (i.e. tensor) polarizations of gravitational
waves which arises from General Relativity in the TT gauge, see previous Subsection,  while the term
$\Phi(t+z)e_{\mu\nu}^{(s)}$ is the extension of the TT gauge to the scalar-tensor case.

For a purely scalar GW the metric perturbation \rfr{eq: perturbazione totale}
reduces to \cite{Essay, Cafaro, Capozziello}

\begin{equation}
h_{\mu\nu}=\Phi e_{\mu\nu}^{(s)},\label{eq: perturbazione scalare}\end{equation}

and the correspondent line element is \cite{Essay, Cafaro, Capozziello}
\begin{equation}
ds^{2}=dt^{2}-dz^{2}-(1+\Phi)dx^{2}-(1+\Phi)dy^{2},\label{eq: metrica TT scalare}\end{equation}
with $\Phi=\Phi_{0}e^{i\omega(t+z)}.$

Again, the wordlines $x,y,z=const.$ are timelike geodesics representing the histories of free test masses, see the analogy with tensor waves in previous Subsection.
In this case, the coordinate transformation $x^{\alpha}=x^{\alpha}(x_{tt}^{\beta})$ from the TT coordinates to the frame of the local observer is \cite{Cafaro}

\begin{equation}
\begin{array}{c}
t=t_{tt}+\frac{1}{4}(x_{tt}^{2}-y_{tt}^{2})\dot{\Phi}\\
\\x=x_{tt}+\frac{1}{2}x_{tt}\Phi+\frac{1}{2}x_{tt}z_{tt}\dot{\Phi}\\
\\y=y_{tt}+\frac{1}{2}y_{tt}\Phi+\frac{1}{2}y_{tt}z_{tt}\dot{\Phi}\\
\\z=z_{tt}-\frac{1}{4}(x_{tt}^{2}-y_{tt}^{2})\dot{\Phi},\end{array}\label{eq: trasf. coord. scal}\end{equation}

where it is $\dot{\Phi}\doteq\frac{\partial\Phi}{\partial t}$ , see previous Subsection and \cite{Cafaro}.

Now, if one considers a free mass riding on a timelike geodesic ($x=l_{1}$, $y=l_{2},$ $z=l_{3}$),   \rfr{eq: trasf. coord. scal} define the
motion of this mass due to the scalar GW with respect to the introduced frame of the local observer. Thus,  one gets

\begin{equation}
\begin{array}{c}
x(t)=l_{1}+\frac{1}{2}l_{1}\Phi(t)+\frac{1}{2}l_{1}l_{3}\dot{\Phi}(t)\\
\\y(t)=l_{2}+\frac{1}{2}l_{2}\Phi(t)+\frac{1}{2}l_{2}l_{3}\dot{\Phi}(t)\\
\\z(t)=l_{3}-\frac{1}{4[}(l_{1}^{2}-l_{2}^{2})\dot{\Phi}(t).\end{array}\label{eq: Grishuk 0 scal}\end{equation}

In absence of scalar GWs the position of the mass is $(l_{1},l_{2},l_{3}).$ Again, the effect of the scalar GW is to drive the mass to have oscillations.
Thus, in general, from   \rfr{eq: Grishuk 0 scal} all three components of motion are present.

Neglecting the terms with $\dot{\Phi}$ in   \rfr{eq: Grishuk 0 scal}, the \emph{traditional} equations for the mass motion due to the scalar GW are obtained \cite{Cafaro}

\begin{equation}
\begin{array}{c}
x(t)=l_{1}+\frac{1}{2}l_{1}\Phi(t)\\
\\y(t)=l_{2}+\frac{1}{2}l_{2}\Phi(t)\\
\\z(t)=l_{3}.\end{array}\label{eq: traditional scal}\end{equation}
This is the analogous of the \emph{electric} component of motion in electrodynamics (see previous Subsection), while equations

\begin{equation}
\begin{array}{c}
x(t)=l_{1}+\frac{1}{2}l_{1}l_{3}\dot{\Phi}(t)\\
\\y(t)=l_{2}+\frac{1}{2}l_{2}l_{3}\dot{\Phi}(t)\\
\\z(t)=l_{3}-\frac{1}{4}(l_{1}^{2}-l_{2}^{2})\dot{\Phi}(t),\end{array}\label{eq: news scal}\end{equation}

are the analogue of the \emph{magnetic} component of motion.

Thus, the \emph{magnetic} component becomes important when the frequency of the wave increases in this case too, but only in the low-frequency regime in analogy with the standard tensor case.

Even in this \emph{scalar} case, one could think that the presence of this \emph{magnetic} component is a \emph{frame artefact} due to the transformation \rfr{eq: trasf. coord. scal}, but now we show that   \rfr{eq: news scal} can be  directly obtained from the geodesic deviation equation too, proving that the \emph{magnetic} components have a real physical significance.

Following \cite{Cafaro}, let us focus the attention on the geodesic deviation extended to second order approximation.
The derivation of the geodesic deviation equations is usually based on a two parameter family of timelike geodesic $x^{\alpha}(\tau,r)$. In general and in the lowest approximation, the geodesic deviation equations are given by \cite{Cafaro, Capozziello}

\begin{equation}
\frac{D^{2}n^{\delta}}{d\tau^{2}}=R_{\alpha\beta\gamma}^{\quad\delta}u^{\alpha}u^{\gamma}n^{\beta}.\label{eq: def geo}\end{equation}

The vector $u^{\alpha}$ is the unit tangent vector to the geodesic and $n^{\alpha}$ is the \emph{separation} vector between two nearby
geodesics

\begin{equation}
u^{\alpha}(\tau,r)=\frac{\partial x^{\alpha}}{\partial\tau}\mid_{r={\rm const.}},\label{eq: u}\end{equation}
\begin{equation}
n^{\alpha}(\tau,r)=\frac{\partial x^{\alpha}}{\partial r}\mid_{t={\rm const.}}.\label{eq: n}\end{equation}

It is also assumed that the central geodesic line corresponds to $r=0$ while the second nearby geodesic corresponds to $r=r_{0}$ \cite{Cafaro}. Then,
$R_{\alpha\beta\gamma}^{\quad\delta}$ in \rfr{eq: def geo} is the curvature tensor calculated along the central geodesic and $\frac{D}{d\tau}$ the covariant derivative calculated along that line \cite{Cafaro}.

To discuss the \emph{magnetic} component of motion in the field of a scalar GW, we need the geodesic deviation equations extended to the next approximation. These equations have been obtained in \cite{Cafaro}. Let us introduce the closely related vector $w^{\alpha}$

\begin{equation}
w^{\alpha}=\frac{Dn^{\alpha}}{dr}=n_{;\beta}^{\alpha}n^{\beta}=\frac{\partial^{2}x^{\alpha}}{\partial r^{2}}+\Gamma_{\beta\gamma}^{\alpha}u^{\beta}u^{\gamma}.\label{eq: w}\end{equation}

This vector obeys the equations \cite{Cafaro}

\begin{equation}
\frac{D^{2}w^{\delta}}{d\tau^{2}}=R_{\alpha\beta\gamma}^{\quad\delta}u^{\alpha}u^{\gamma}w^{\beta}+(R_{\alpha\beta\gamma;\epsilon}^{\quad\delta}-R_{c\epsilon\alpha;\beta}^{\quad\delta})u^{\alpha}u^{\beta}u^{\gamma}u^{\epsilon}+4R_{\alpha\beta\gamma}^{\quad\delta}u^{\beta}\frac{Dn^{\alpha}}{d\tau}n^{\gamma}.\label{eq: geo estese}\end{equation}

Defining the vector

\begin{equation}
N^{\alpha}\doteq r_{0}n^{\alpha}+\frac{1}{2}r_{0}^{2}w^{\alpha},\label{eq: N}\end{equation}

\rfr{eq: def geo} and \rfr{eq: w} can be combined
obtaining \cite{Cafaro}

\begin{equation}
\frac{D^{2}N^{\delta}}{d\tau^{2}}=R_{\alpha\beta\gamma}^{\quad\delta}u^{\alpha}u^{\gamma}N^{\beta}+(R_{\alpha\beta\gamma;\epsilon}^{\quad\delta}-R_{\gamma\epsilon\alpha;\beta}^{\quad\delta})u^{\alpha}u^{\beta}N^{\gamma}N^{\epsilon}+2R_{\alpha\beta\gamma}^{\quad\delta}u^{\beta}\frac{DN^{\alpha}}{d\tau}N^{\gamma}+{\mathcal{O}}(r_{0}^{3}).\label{eq: geo estesissime}\end{equation}

Thus, it is possible writing the expansion of $x^{\alpha}(\tau,r_{0})$ in terms of $N^{\alpha}$ \cite{Cafaro}

\begin{equation}
x^{\alpha}(\tau,r_{0})=x^{\alpha}(\tau,0)+N^{\alpha}-\Gamma_{\beta\gamma}^{\alpha}N^{\beta}N^{\gamma}+{\mathcal{O}}(r_{0}^{3}).\label{eq: exp}\end{equation}

This formula shows that in the frame of the local observer (in which it is $\Gamma_{\beta\gamma}^{\alpha}=0$ along the central geodesic line \cite{Cafaro}) the spatial components of $N^{\alpha}$ will directly give the time-dependent position of the nearby test mass.
According to    \rfr{eq: geo estesissime}, these positions include the next-order corrections, as compared with solutions to    \rfr{eq: def geo}.

Now, let us specialize to the  scalar GW metric \rfr{eq: metrica TT scalare}. We take into account only the linear perturbations in terms of the scalar GW amplitude $\Phi.$ The first test mass is described by the central timelike geodesic $x^{i}(t)=0$. The correspondent tangent vector is $u^{\alpha}=(1,0,0,0)$. The second test mass is situated in the unperturbed position $x^{i}(0)=l^{i}$ having zero unperturbed velocity \cite{Cafaro}.
We assume that the frame of the local observer is located along the central geodesic. The goal is to find the trajectory of the second test mass using the geodesic deviation equation \rfr{eq: geo estesissime}.
The deviation vector can be written like \cite{Cafaro}
\begin{equation}
N^{i}(t)=l^{i}+\delta l^{i}(t)\label{eq: N2}\end{equation}
where the variation in distance $\delta l^{i}(t)$ is caused by the scalar GW. Using the frame of the local observer, one can replace all the covariant derivatives in    \rfr{eq: geo estesissime} by ordinary derivatives \cite{Cafaro, Gravitation}. In the lowest approximation    \rfr{eq: geo estesissime}
reduces to    \rfr{eq: def geo} and specializes to
\begin{equation}
\frac{d^{2}\delta l^{i}(t)}{dt^{2}}=-\frac{1}{2}l^{j}\frac{\partial^{2}}{\delta t^{2}}\Phi\delta_{j}^{i}=\frac{1}{2}\omega^{2}l^{j}\Phi e_{\textrm{ $j$ }}^{(s)i}\label{eq: delta 1}\end{equation}
in the field of a scalar GW \rfr{eq: metrica TT scalare}. The relevant solution to  \rfr{eq: delta 1} coincides exactly with the
usual \emph{electric} part of the motion given by equation \rfr{eq: traditional}. As we want to identify the \emph{magnetic} part of the gravitational force arising from a scalar GW, all the terms in \rfr{eq: geo estesissime} have to be considered. Since $\frac{DN^{a}}{d\tau}$ is of the order of $\Phi$, the third term of  \rfr{eq: geo estesissime} is of the order of $\Phi^{2}$ and can be neglected. Working out the derivatives of the curvature tensor and substituting them into equation \rfr{eq: geo estesissime} specialized in the field of a scalar GW \rfr{eq: metrica TT scalare}, the accurate equations of motions read
\begin{equation}
\frac{d^{2}\delta l^{i}(t)}{dt^{2}}=\frac{1}{2}\omega^{2}l^{j}\Phi e_{\textrm{ $j$ }}^{(s)i}-\frac{1}{2}\omega^{2}l^{k}l^{l}(k_{l}\delta^{ij}+\frac{1}{2}k^{i})\delta_{l}^{j}\Phi e_{kj}^{(s)}.\label{eq: delta 2}\end{equation}
In this equation, which clearly exhibits two contributions, the second term is responsible for the \emph{magnetic} component of motion and can be interpreted as the gravitational analogue of the magnetic part of the Lorentz force (see also the analogy for ordinary tensor waves in previous Subsection).

\subsection {Variation of distances between test masses and response of interferometers for the scalar \emph{magnetic} component}

It was already recalled that the previous descriptions in the frame of the local observer are as close as possible to the description of laboratory physics. As all the questions concerning test masses positions have been analysed, now it is possible discussing the variation of distances. We are interested in the distance between the central particle, located at coordinate origin, and the particle located, on average, at some position $(l_{1},l_{2},l_{3})$. This model represents the situation of the beam - splitter and one mirror of an interferometer \cite{Cafaro, Capozziello}. In the frame of the local observer the line element is given by equation \cite{Gravitation}

\begin{equation}
ds^{2}=-(dx^{0})^{2}+\delta_{ij}dx^{i}dx^{j}+{\mathcal{O}}(|x^{j}|^{2})dx^{\alpha}dx^{\beta};\label{eq: metrica local lorentz}\end{equation}

which gives the Galileian distance

\begin{equation}
d(t)=\sqrt{x^{2}+y^{2}+z^{2}}+{\mathcal{O}}([\Phi l(\omega l)]{}^{2}).\label{eq: metrica local lorentz 2}\end{equation}

\Rfr{eq: metrica local lorentz 2} is accurate for terms of the order of $\Phi l$ and $\Phi l^{2}\omega$ inclusive, while the terms quadratic in $\Phi$ are neglected. Putting

\begin{equation}
\begin{array}{c}
x=l_{1}+\delta x\\
y=l_{2}+\delta y\\
z=l_{3}+\delta z,\end{array}\label{eq: delta xyz}\end{equation}

we get \cite{Cafaro}

\begin{equation}
d(t)=l+\frac{1}{l}(l_{1}\delta x+l_{2}\delta y+l_{3}\delta z)\label{eq: distanza}\end{equation}
and, using the time dependent positions \rfr{eq: news scal}, the
distance $d(t)$ is obtained with the required approximation (i.e. $\omega l\ll1$)

\begin{equation}
d(t)=l+\frac{1}{2l}(l_{1}^{2}-l_{2}^{2})\Phi(\omega t)-\frac{1}{4l}\omega l_{3}(l_{1}^{2}-l_{2}^{2})\Phi\left(\omega t-\frac{\pi}{2}\right).\label{eq: distanza 2}\end{equation}

Clearly, the first correction to $l$ is due to the \emph{electric} contribution, while the second correction to $l$ is due to the \emph{magnetic}
contribution.

Now, let us compute the response of a laser interferometer. To compute the response function for an arbitrary propagating direction of the scalar GW one recalls that the arms of the interferometer are in the $\overrightarrow{u}$ and $\overrightarrow{v}$ directions, while the $x,y,z$ frame is adapted to the propagating scalar GW. Then, once again, the spatial rotation of the coordinate \rfr{eq: rotazione magn} has to be performed.

In this way the scalar GW is propagating from an arbitrary direction $\overrightarrow{r}$ to the interferometer (see Figure 2).

At this point, one recalls that the response function is given by

\begin{equation}
\delta d(t)\doteq d_{u}(t)-d_{v}(t),\label{eq: funzione risposta}\end{equation}

where $d_{u}(t)$ and $d_{v}(t)$ are the distances in the $u$ and $v$ direction,  and, using equations \rfr{eq: distanza 2}, \rfr{eq: rotazione magn}, and \rfr{eq: funzione risposta} it is
\begin{equation}
\delta d(t)
=
-\Phi(t) l \sin^{2}\theta \cos2\phi
+
\Phi(t) \omega l^{2} \frac{1}{4} \cos\theta
\left\{
\left[
\left(
\frac{1+\sin^{2}\theta}{2}
\right)
+
\sin^{2}\theta\sin2\phi
\right](\cos\phi-\sin\phi)
\right \}.\label{eq: funzione risposta 2}\end{equation}
In this equation the first term is due to the \emph{electric} contribution, while the second term is due to the \emph{magnetic} contribution \cite{Cafaro}.
The function
\begin{equation}
\omega l\frac{1}{4}\cos\theta\left\{\left[\left(\frac{1+\sin^{2}\theta}{2}\right)+\sin^{2}\theta\sin2\phi\right](\cos\phi-\sin\phi)\right\}\label{eq: pattern}\end{equation}
represents the so-called \emph{angular pattern} \cite{Cafaro} of interferometers for the \emph{magnetic} contribution. The frequency-dependence in this angular pattern renders the \emph{magnetic} component important in the high-frequency portion of the interferometers sensitivity band.
Its value is shown in Figure 5 for the LIGO interferometer for the frequency $f=8000$ Hz, which falls in such a high-frequency portion.

\begin{figure}
\includegraphics{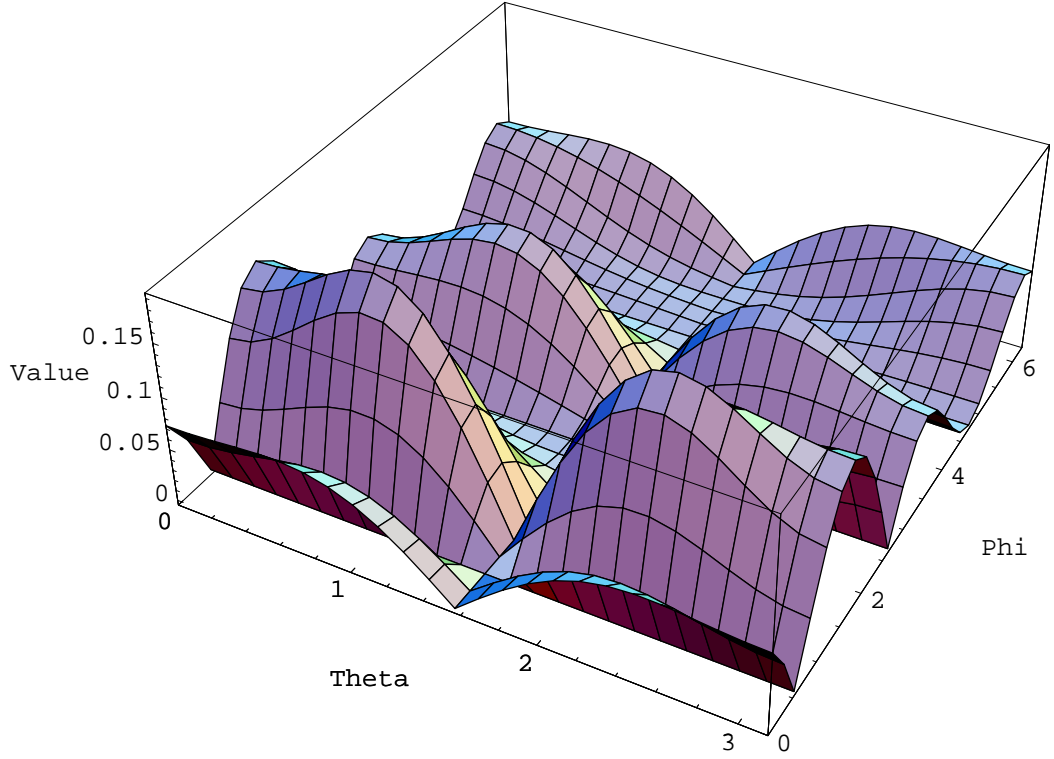}\lb{figuracinque}

\caption{the angular dependence of the total response function of the LIGO
interferometer to the magnetic component of a SGW for $f=8000$ Hz}
\end{figure}

Again, from the Figure, it looks clear that if one neglects the \emph{magnetic} contribution, approximately $15\%$ of currently observable signal could,
in principle, be lost \cite{Cafaro}.

\section{Conclusion remarks}

After extensively reviewing general relativistic gravitomagnetism, both historically and phenomenologically, the so-called \emph{magnetic} components of GWs have been reviewed in detail. Such  components have to be taken into account in the context of the total response functions of interferometers for GWs propagating from arbitrary directions.
Following the more recent approaches of this important issue, the analysis of such \emph{magnetic} components has been reviewed in both of standard GTR and Scalar Tensor Gravity. Thus, it has been shown in detail that such a \emph{magnetic} component becomes particularly important in the high-frequency portion of the range of ground based interferometers for GWs which arises from the two different theories of gravity.
the reviewed results have shown that if one neglects the \emph{magnetic} contribution to the gravitational field of a GW, approximately $15\%$ of the potential observable signal could, in principle, be lost.

\end{document}